\documentclass[reprint,amsmath,nofootinbib,amssymb,aps,showpacs,prc]{revtex4-1}
\usepackage{graphicx}
\usepackage{subfigure}
\usepackage{multirow}
\usepackage[colorlinks,
linkcolor=blue,
anchorcolor=blue,
citecolor=blue]{hyperref}
\usepackage[all]{hypcap}
\usepackage{epstopdf}
\usepackage{caption}
\usepackage{color}
\usepackage{bm}

\begin{document}


\title{Practical considerations for measuring global spin alignment of vector mesons in relativistic heavy ion collisions }

\author{A. H. Tang$^1$, B. Tu$^{2}$ and C. S. Zhou$^{3,4}$}
\affiliation{\mbox{$^1$Brookhaven National Laboratory, Upton, New York 11973 USA}\\
$^2$Key Laboratory of Quark and Lepton Physics (MOE) and Institute of Particle Physics,\\ \mbox{Central China Normal University, Wuhan, Hubei 430079, China}\\
\mbox{$^3$Shanghai Institute of Applied Physics, Chinese Academy of Sciences, Shanghai 201800 P.R. China}\\
\mbox{$^4$ University of Chinese Academy of Sciences, Beijing 100049 P.R. China}
}

\begin{abstract}
Global spin alignment of vector mesons is a sensitive probe of system vorticity and particle production mechanism in relativistic heavy ion collisions. The measurement of global spin alignment is gaining increasing interest and deserves careful considerations. In this paper, we lay out a few practical issues that need to be taken care of when measuring global spin alignment of vector mesons. They are, the correction for event plane resolution, reconciling measurements made with different event planes, the correction for the effect of finite acceptance in pseudorapidity, and the consideration for measuring the azimuthal angle dependence. Insights and methodologies offered in this paper will help experiments to measure the global spin alignment properly and accurately.  
\end{abstract}


\maketitle


\section{Introduction} \label{sec:intro}
\vspace{-0.08cm}
In non-central relativistic heavy-ion collisions the initial, global angular momenta can be as large as of the order of 1000$\hbar$. A significant fraction of the global angular momentum ($\hat{L}$) is deposited in the interaction region, producing a finite gradient of the longitudinal momentum of produced partons.  Such gradient is in the direction perpendicular to the incident nuclei, and it generates vorticity for partons, along the direction of global angular momentum. The vorticity of partons is eventually transferred to spin degree of freedom of final-state hadrons~\cite{Liang:2004ph,Voloshin:2004,Liang:2004xn,Liang:2007ma,Betz:2007kg,Gao:2007bc,Becattini:2013vja}, with the spin direction on average aligning globally with the direction of $\hat{L}$. Although in general this picture is consistent with first principle expectations, how the vorticity field evolves in the system and how spin transports within fluid is not yet fully understood~\cite{Montenegro:2017rbu,Florkowski:2017ruc}. The comprehensive study of global polarization of hyperons and spin alignment of vector mesons can help us to probe the vorticity field~\cite{Becattini:2007sr,Pang:2016igs,Becattini:2015ska, Teryaev:2015gxa, Jiang:2016woz, Li:2017slc} and, from the spin transport aspect, understand particle production mechanisms during hardronization~\cite{Liang:2004xn}. For those reasons this topic is gaining increasing interest~\cite{Abelev:2007zk,Abelev:2008ag,STAR:2017ckg,Mohanty:2017sqm,Singh:2018uad,Voloshin:2017sqm,Becattini:2017sqm}, and in particular it is recently demonstrated that the BNL Relativistic Heavy Ion Collider has produced the most vortical fluid with least viscosity~\cite{STAR:2017ckg}. 

Although global polarization of hyperons and spin alignment of vector mesons are both caused by the vorticity, they bear different sensitivity to it. For sin-1 vector mesons, their daughter's polar angle distribution is an even function (see more below)  as opposite to an odd function~\cite{Abelev:2007zk} for hyperons with a spin of 1/2. Because of that, the global polarization is sensitive to the sign of vorticity field, while the global spin alignment is only sensitive to its strength. This feature does not necessarily mean a disadvantage for the spin alignment measurement, as for spin alignment there is no local cancellation associated with odd function when integrating over time and phase space. However, it does mean that the reconstruction procedure for spin alignment is very different to that for global spin alignment. In this paper, we lay out a few practical considerations when analyzing the global spin alignment of vector mesons. They are, the derivation of correction for finite event plane resolution, the relation between results obtained with the first and second order event plane, the procedure to correct for finite acceptance for experimental setup with cylindrical symmetry, and the consideration for extracting the spin alignment measurement as a function of azimuthal angle w.r.t the reaction plane and how to correct for the smearing in azimuth due to finite event plane resolution.

\section{Correction for finite event plane resolution} \label{sec:resolution}

For spin-1 vector meson, the spin alignment can be described by a $3\times3$ spin density matrix $\rho$ with unit trace~\cite{Schilling:1969um}. The deviation of the diagonal elements of $\rho_{mm}$ $(m=-1,0,1)$ from $1/3$ signals net spin alignment.  Out of three diagonal elements, $\rho_{11}$ and $\rho_{-1-1}$ are degenerate, and $\rho_{00}$ is independent of the other two. We consider a two-body decay of a spin-1 vector meson. Without loss of generality, let's take $\phi \rightarrow K^+ + K^-$ as an example, the angular distribution of one of the decay products (say, $K^+$) can be written as:
\begin{eqnarray}
f(\theta^*)=\frac{dN}{d(\mathrm{cos}\theta^*)} \propto (1-\rho_{00}) + (3 \rho_{00} -1) \mathrm{cos}^2\theta^*,
\label{eq:rho_00}
\end{eqnarray}
where $\theta^*$ is is the angle between the quantization axis and the momentum direction of $K^+$ in the rest frame of $\phi$ particle. 

\begin{figure}[h]       
\centering
\makebox[1cm]{\includegraphics[width=0.45 \textwidth]{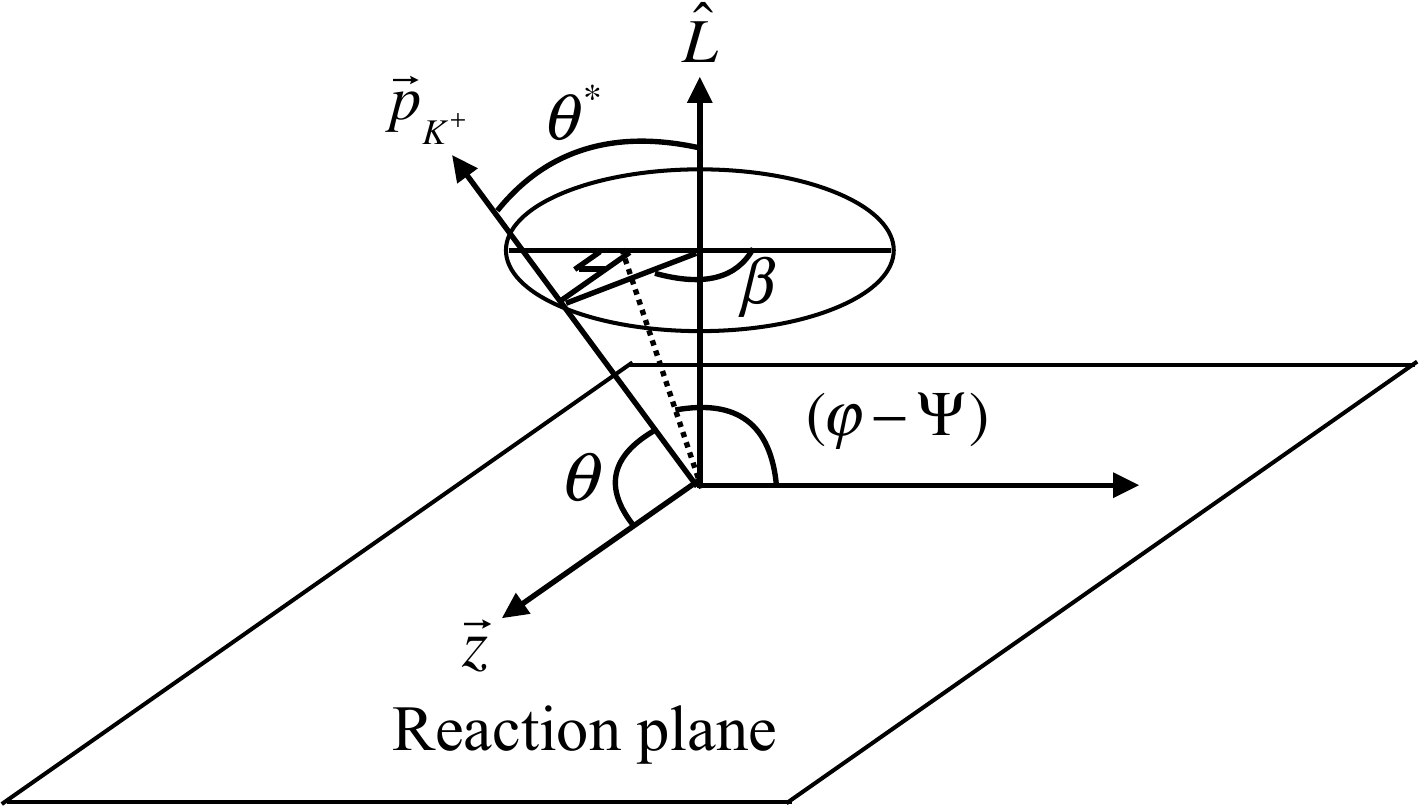}}
\caption{The definition of angles. $\vec{p}_{K^+}$ is the momentum of the daughter particle $K^+$ in the rest frame of $\phi$ meson. 
}
  \label{fig:angles}
\end{figure}

The relevant angles are illustrated in Fig.~\ref{fig:angles}. To study the global spin alignment, the quantization axis is set to be in the direction of $\hat{L}$. The beam line ($\hat{z}$-axis) and the the line (not shown) connecting the centers of two colliding nuclei defines the reaction plane (RP),  which intersects the plane that is transverse to beam line (transverse plane) at an azimuthal angle $\Psi$. In $\phi$ particle's rest frame and with respect to $\hat{L}$ , $\theta^*$ is the polar angle of $K^+$ momentum and $\beta$ is its azimuthal angle. In the same frame, $\theta$ is the angle between $K^+$ momentum and $\hat{z}$-axis and $(\varphi-\Psi)$ is its azimuthal angle w.r.t RP.  These angles are related to each other by the relation :
\begin{eqnarray}
\begin{aligned}
\mathrm{cos}\theta^* & = \mathrm{sin}{\theta}\mathrm{sin}(\varphi-\Psi) , \\
\mathrm{cos}\theta & = \mathrm{sin}{\theta^*}\mathrm{sin}\beta .
\end{aligned}
\label{eq:trigonometric1}
\end{eqnarray}

In experiment the reaction plane angle $\Psi$ cannot be known \textit{a priori}, but it can be estimated by the study of correlation between produced particles~\cite{Poskanzer:1998yz}. Usually the estimated reaction plane (event plane, EP) has finite resolution, its distribution is centered around the true RP with finite width. Because the $\hat{L}$ direction is by definition the normal of $\Psi$, the smearing of $\Psi$ is equivalent to the smearing of  $\hat{L}$ , and in general will artificially decrease the observed value of $\rho_{00}$. To recover the real $\rho_{00}$ from the observed one, an event plane resolution correction is needed. In the past such correction factor was obtained with a complicated simulation procedure~\cite{Abelev:2008ag,Mohanty:2017sqm,Singh:2018uad}, in this paper we derive an analytic formula for a straight forward correction. 

We start with a general form for the daughter's angular distribution for spin-1 particles~\cite{Schilling:1969um}
\begin{eqnarray}
\begin{aligned}
\frac{ d \mathrm{ N } } {d\mathrm{cos}\theta^*d\beta} \propto & 1+A\mathrm{cos}^2\theta^*+B\mathrm{sin}^2\theta^*\mathrm{cos}2\beta \\
& +C\mathrm{sin}2\theta^*\mathrm{cos}\beta,
\label{eq:distrib}
\end{aligned}
\end{eqnarray}
where 
\begin{eqnarray}
A=\frac{3\rho_{00}-1}{1-\rho_{00}} .
\label{eq:Arho_00}
\end{eqnarray}
The EP $\Psi'$ of each event can be viewed as rotated  RP by an angle $\Delta$, 
\begin{eqnarray}
\Psi' = \Psi + \Delta .
\end{eqnarray}
Note that the distribution of $\Delta$ over many events is an even function centered at zero.  Thus 
\begin{eqnarray}
\left\langle{\mathrm{sin}2\Delta}\right\rangle = 0,  \left\langle{\mathrm{cos}2\Delta}\right\rangle \equiv R \mathrm {\,(EP\, resolution)} .
\label{eq:Delta}
\end{eqnarray}
Under the rotation from RP to EP,  $\theta^*\rightarrow\theta'^*$,  and $\beta\rightarrow\beta'$, and $\theta$ remains unchanged. After the rotation, we have similar trigonometric relation between $\theta'^*$, $\beta'$ and $\theta$ 
\begin{eqnarray}
\begin{aligned}
\mathrm{cos}\theta'^* & = \mathrm{sin}{\theta}\mathrm{sin}(\varphi-\Psi') ,\\
\mathrm{cos}\theta & = \mathrm{sin}{\theta'^*}\mathrm{sin}\beta' .
\end{aligned}
\label{eq:trigonometric2}
\end{eqnarray}
The daughter's distribution under the rotated frame can be written in the same form as Eq.~\eqref{eq:distrib}  :
\begin{eqnarray}
\begin{aligned}
\frac{ d \mathrm{ N } } {d\mathrm{cos}\theta'^*d\beta'} \propto & 1+A'\mathrm{cos}^2\theta'^*+B'\mathrm{sin}^2\theta'^*\mathrm{cos}2\beta' \\ 
& +C'\mathrm{sin}2\theta'^*\mathrm{cos}\beta',
\label{eq:distribRotate}
\end{aligned}
\end{eqnarray}
where the coefficient $A'$ is related to the observed spin alignment $\rho^{obs}_{00}$ by 
\begin{eqnarray}
A'=\frac{3\rho^{obs}_{00}-1}{1-\rho^{obs}_{00}} .
\label{eq:Arho_00_2}
\end{eqnarray}
Taking Eq.~\eqref{eq:trigonometric1} and Eq.~\eqref{eq:trigonometric2}, and keeping in mind that when averaging over many events at a later time, terms containing $\mathrm{sin}\Delta$ will disappear and terms containing $\mathrm{cos}2\Delta$ will becomes the resolution (Eq.~\eqref{eq:Delta}), we can express terms in the original distribution Eq.~\eqref{eq:distrib} with terms in the rotated frame,
\begin{eqnarray}
\begin{aligned}
&{\mathrm{cos}^2\theta^*}
=  \mathrm{sin}^2\theta \, \mathrm{sin}^2(\varphi-\Psi'+\Delta) \\
= & \frac{1 \!-\! R}{4}
 \!+\! \frac{1 \!+\! 3R}{4}\mathrm{cos}^2\theta'^*
 \!+\! \frac{1 \!-\! R}{4}\mathrm{sin}^2\theta'^*\mathrm{cos}2\beta' ,
\end{aligned}
\end{eqnarray}
\begin{eqnarray}
\begin{aligned}
&
\mathrm{sin}^2\theta^*\mathrm{cos}2\beta =  1 - \mathrm{cos}^2\theta^*
- 2\mathrm{sin}^2\theta'^*\mathrm{sin}^2\beta' \\
 = & \!-\! \frac{1 \!-\! R}{4}
 \!+\! \frac{3 \!-\! 3R}{4}\mathrm{cos}^2\theta'^*
 \!+\! \frac{3 \!+\! R}{4}\mathrm{sin}^2\theta'^*\mathrm{cos}2\beta' ,
\end{aligned}
\end{eqnarray}
\begin{eqnarray}
\begin{aligned}
\mathrm{sin}2\theta^*\mathrm{cos}\beta &= \mathrm{sin}^2\theta \, {\mathrm{sin}2(\varphi-\Psi'+\Delta)} \\
&= R\cdot\mathrm{sin}2\theta'^*\mathrm{cos}\beta' .
\end{aligned}
\end{eqnarray}
Inserting back into Eq.~\eqref{eq:distrib} and after reorganization, we establish the relation between coefficients for distribution in the original frame and the rotated one as :
\begin{eqnarray}
\begin{pmatrix} 1 \\ A \\ B \\ C \end{pmatrix}
\rightarrow
\begin{pmatrix} 1 \\ A' \\ B' \\ C' \end{pmatrix}
=
\begin{pmatrix} 1 \\
\frac{A(1+3R)+B(3-3R)}{4+A(1-R)+B(-1+R)} \\
\frac{A(1-R)+B(3+R)}{4+A(1-R)+B(-1+R)} \\
\frac{C\cdot 4R}{4+A(1-R)+B(-1+R)} .
\end{pmatrix}
\label{eq:matrixABC}
 \end{eqnarray}

The distribution with the form of Eq.~\eqref{eq:rho_00} is obtained by integrating the general distribution Eq.~\eqref{eq:distrib} over $\beta$, which is equivalent of letting $B=0$ and $C=0$. With that we have
\begin{eqnarray}
A'=\frac{A(1+3R)}{4+A(1-R)} ,
\end{eqnarray}
and expressing $A$ and $A'$ in terms of $\rho_{00}$ and $\rho^{obs}_{00}$, by Eq.~\eqref{eq:Arho_00} and Eq.~\eqref{eq:Arho_00_2} respectively,  we obtain
\begin{eqnarray}
\boxed{\rho_{00} - \frac{1}{3} = \frac{4}{1+3R}(\rho^{obs}_{00} - \frac{1}{3}) } .
\label{eq:resolution}
\end{eqnarray}
which is the formula for the resolution correction for $\rho_{00}$ .  

Eq.~\eqref{eq:resolution} has interesting implications. Noting that the deviation of $\rho_{00}$ from 1/3 quantifies how a shape deviates from a ball-like shape (no spin alignment),  this  equation tells us that in general even when the resolution for RP is zero ($R=0$ and EP angle is random w.r.t $z$-axis), one should not expect $\rho^{obs}_{00}$ to be at 1/3. This sounds striking at first but it is understandable. It means that in general an object cannot be rotated into a ball around a fixed axis, unless the object itself is a ball. 

\begin{figure}[h]       
\centering
\makebox[1cm]{\includegraphics[width=0.5 \textwidth]{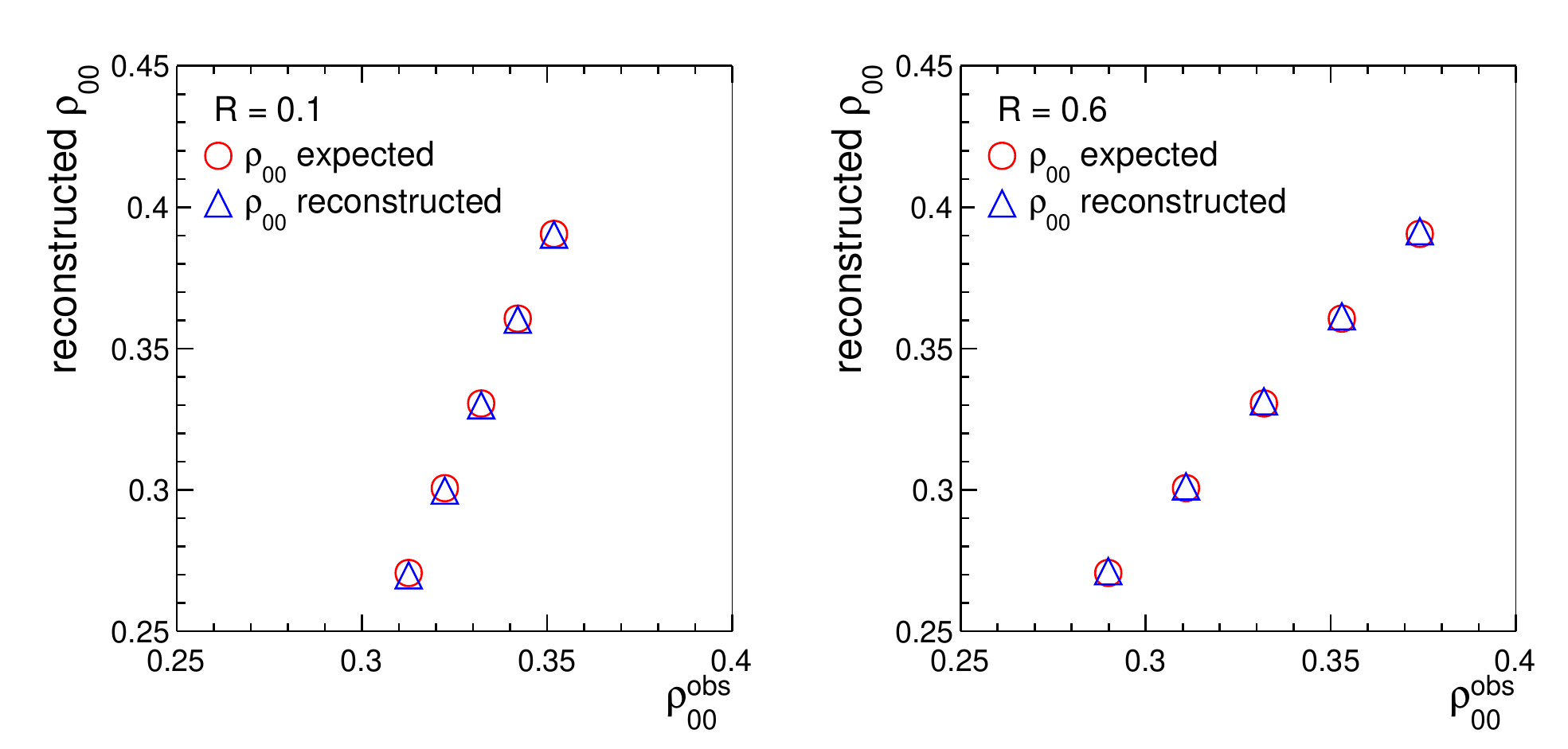}}
\caption{The simulation of EP resolution correction, for a relatively small resolution (left, $R = 0.1$) and large resolution (right, $R = 0.6$). Red symbols are $\rho_{00}$ extracted from fitting actual $\mathrm{cos}\theta^*$ distributions (expected values), and $\rho_{00}^{\mathrm{real}}$ for blue symbols are reconstructed with known $\rho_{00}^{obs}$ and $R$. }
  \label{fig:resolutionSim}
\end{figure}

The correction formula Eq.~\eqref{eq:resolution} has been tested with simulation, as shown in Fig.~\ref{fig:resolutionSim}. In this simulation, particles are generated according to a probability density function of Eq.~\eqref{eq:rho_00} with a RP, then a set of $\Delta$ with respect to RP are generated with the following probability density function~\cite{Voloshin:1994mz}
\begin{eqnarray}
\begin{aligned}
 \mathcal{P}(\Delta) = \frac{1}{2\pi} & \left[ e^{-\frac{\chi^{2}}{2}}+\sqrt{\frac{\pi}{2}}\chi\mathrm{cos}(\Delta) e^{-\frac{\chi^{2}\mathrm{sin}^{2}(\Delta)}{2}} \right. \\
& \left. \times (1+\mathrm{erf}(\chi\mathrm{cos}\frac{\Delta}{\sqrt{2}})) \right] ,
\label{eq:Detla_Psi_dis}
\end{aligned}
\end{eqnarray}
where $\chi$ can be determined with known EP resolution $R$~\cite{Poskanzer:1998yz}.
The particle's $\mathrm{cos}\theta^*$ distribution with EP is used to extract the $\rho_{00}^{obs}$. The red circles are for $\rho_{00}$ reconstructed by fitting the actual distribution, with $\rho_{00}^{obs}$ and $\rho_{00}^{\mathrm{real}}$ extracted w.r.t EP and RP, respectively. The blue symbols are calculated $\rho_{00}^{\mathrm{real}}$ with Eq.~\eqref{eq:resolution}, with known $\rho_{00}^{obs}$ and resolution $R$ as input. They agree with each other very well, indicating that the formula for the resolution correction works as expected.

A different method of obtaining $\rho_{00}$ is given in~\cite{Voloshin:2017sqm} by calculating the mean value of $\cos{2(\varphi-\Psi)}$:
\begin{eqnarray}
\begin{aligned}
\rho_{00} - \frac{1}{3} &= -\frac{4}{3}\left\langle\cos{2(\varphi-\Psi)}\right\rangle \\
&=-\frac{4}{3}\left\langle\cos{2(\varphi-\Psi')}\right\rangle/R .
\end{aligned}
\label{eq:Sergeiresolution}
\end{eqnarray}
Unlike elliptic flow analysis~\cite{Poskanzer:1998yz} for which the formula for observed anisotropy is obtained by simply replacing RP with EP,  here in the first line of Eq.~\eqref{eq:Sergeiresolution} replacing RP ($\Psi$) by the EP ($\Psi'$) will not result in the observed $\rho^{obs}_{00}$. Instead, the relation between $\rho^{obs}_{00}$ and $\left\langle\cos{2(\varphi-\Psi')}\right\rangle$ is given as :
\begin{eqnarray}
\rho^{obs}_{00} - \frac{1}{3} = -\frac{1+3R}{3R}\left\langle\cos{2(\varphi-\Psi')}\right\rangle .
\label{eq:relationBetweenSergeiAndus}
\end{eqnarray}
Note that calculating $\left\langle\cos{2(\varphi-\Psi')}\right\rangle$ does not gain straightforward insight into the $\rho^{obs}_{00}$ obtained w.r.t random EP. This can be seen as, in Eq.~\eqref{eq:relationBetweenSergeiAndus} when $R\rightarrow 0$ (i.e. random event plane), $\left\langle\cos{2(\varphi-\Psi_{EP})}\right\rangle\rightarrow 0$, and the right hand side of Eq.~\eqref{eq:relationBetweenSergeiAndus} becomes a $\frac{0}{0}$ limit for which the value is not immediately clear.

\section{Understand $\rho_{00}$ obtained with different event planes} \label{sec:relation_1st2ndEP}

For experiments like STAR and ALICE, the 1st order EP can be reconstructed by measuring the deflection of neutrons in forward region~\cite{Adams:2005ca,Abelev:2013cva}. $\hat{L}$ direction reconstructed through this procedure is more sensitive to the initial angular momentum, because a large fraction of the initial angular momentum is carried away by spectator neutrons. On the other hand, from time to time $\rho_{00}$ is also measured with the normal of the 2nd order EP as $\hat{L}$, to take advantage of the good EP resolution observed at midrapidity. However, these two planes are usually different to each other, because the deflection of spectator neutrons at forward region is not necessarily in the direction of the principle axis of the ellipsoid shape in momentum space of the system at midrapidity. In general, $\rho_{00}$ reconstructed with Eq.~\eqref{eq:resolution} for the two EPs, each with corresponding $\rho^{obs}_{00}$ and resolution of their own, are different to each other. Below we make a connection between $\rho_{00}$ measured with these two EPs.  Note that here we used the 1st and 2nd order EP as examples but the idea can be easily generalized, for example, to be used in the understanding the difference in results obtained with EPs reconstructed with different detectors or analysis cuts. 

To make a connection between $\rho_{00}$ measured with these two EPs, we need the 2nd order EP's ``resolution" w.r.t the plane that the reconstructed 1st order EP is perturbing around,  
\begin{eqnarray}
R_{21} = \left\langle{\mathrm{cos}2( \Psi_2 - \Psi_{r,1} )}\right\rangle.
\end{eqnarray}
Here $\Psi_2$ is the reconstructed 2nd order EP, and $\Psi_{r,1}$ is the plane around which the reconstructed one ($\Psi_1$) is perturbing with a resolution of $R_1  = \left\langle{\mathrm{cos}2( \Psi_1 - \Psi_{r,1} )}\right\rangle$.
To obtain $R_{21}$, one can factorize the de-correlation between reconstructed $\Psi_1$ and $\Psi_2$ as,
\begin{eqnarray}
\begin{aligned}
D_{12}&\equiv \left\langle{\mathrm{cos}2( \Psi_1 - \Psi_2 )}\right\rangle \\
&= \left\langle{\mathrm{cos}2( \Psi_1 - \Psi_{r,1} + \Psi_{r,1} - \Psi_2 )}\right\rangle \\
&\approx  \left\langle{\mathrm{cos}2( \Psi_1 - \Psi_{r,1} )}\right\rangle \left\langle{\mathrm{cos}2(  \Psi_{r,1} - \Psi_2 )}\right\rangle   \\
&=  R_1 \cdot R_{21}.
\end{aligned}
\end{eqnarray}
Both $D_{12}$ and $R_1$ are measurable quantities thus $R_{21}$ can be calculated as $R_{21} = D_{12} / R_1$. With known $R_{21}$ the $\rho_{00}$ can also be reconstructed by the equation below: 
\begin{eqnarray}
\rho_{00} - \frac{1}{3} = \frac{4}{1+3R_{21}}(\rho^{obs\_2ndEP}_{00} - \frac{1}{3}) .
\label{eq:resolutionWithD}
\end{eqnarray}$\rho_{00}$ obtained with the equation above should, to the first order, be compatible to $\rho_{00}$ directly obtained with 1st EP with Eq.~\eqref{eq:resolution}. Should there be any remaining difference, it can come from finite $B$ in the general distribution Eq.~\eqref{eq:distrib}. A finite $B$ describes the non-uniform distribution in $\beta$, and it has no effect on $\rho_{00}$ in the original frame as we only consider the distribution integrated over $\beta$. But for the same distribution viewed in the rotated frame, it will have an effect on $\rho_{00}^{obs}$. 

\section{Correction for acceptance effect for detectors with cylindrical symmetry} \label{sec:acceptance}
Measuring spin alignment is equivalent to identifying the 3-D shape of daughter's momentum distribution in parent's rest frame, and it is desirable to have the measurement conducted in full phase space. However,  the experimental coverage of phase space is always finite. A finite phase space will introduce a distortion on the shape (and hence $\rho_{00}$), and it is pointed out~\cite{Lan:2017nye} that a finite pseudorapidity ($\eta$) coverage will have significant distortion on the measured $\rho_{00}$. Usually such effect can be corrected for with known efficiencies (as the case in~\cite{Abelev:2008ag}). Here we point out that for experimental setup that has cylindrical symmetry (like STAR and ALICE), there exists a way to quantify this effect independently and correct for it.

We can regard the distribution obtained with finite pseudorapidity acceptance as a convolution of real signal and acceptance effect :
\begin{eqnarray}
\left[ \frac{d\mathrm{N}}{d\cos\theta^* d\beta} \right]_{|\eta|} = \frac{d\mathrm{N}}{d\cos\theta^* d\beta}\times g(\theta^*,\beta).
\end{eqnarray}
Because the acceptance effect is symmetrical w.r.t the z-axis, we can describe it as:
\begin{eqnarray}
\begin{aligned}
g(\theta^*,\beta) = & 1+F^*\cos^2{\theta} \\ 
 \propto &  1+F\cos^2{\theta^*}+F\sin^2{\theta^*}\cos{2\beta} ,
\end{aligned}
\end{eqnarray}
where
\begin{eqnarray}
F=-\frac{F^*}{2+F^*} .
\end{eqnarray}
With the event plane resolution correction and acceptance correction term $g(\theta^*,\beta)$ both considered, we have
\begin{eqnarray}
\begin{aligned}
& \left[ \frac{d\mathrm{N}}{d\cos\theta'^* d\beta'} \right]_{|\eta|} \! \propto \!  (1 \!+\! A'\cos^2{\theta'^*} \!+\! B'\sin^2{\theta'^*}\cos{2\beta'} \\
& \!+\! C'\sin{2\theta'^*}\cos{\beta'}) \! \times \! (1 \!+\! F\cos^2{\theta'^*} \!+\! F\sin^2{\theta'^*}\cos{2\beta'}).
\end{aligned}
\end{eqnarray}
By integrating over $\beta'$, we have
\begin{eqnarray}
\begin{aligned}
\left[ \frac{d\mathrm{N}}{d\cos\theta'^*} \right]_{|\eta|}
\propto &  (1 + \frac{B'F}{2}) + (A'+F)\cos^2{\theta'^*} \\
+ & (A'F-\frac{B'F}{2})\cos^4{\theta'^*},
\end{aligned}
\end{eqnarray}
where $A'$ and $B'$ are related to $A$ and $B$ by Eq.~\eqref{eq:matrixABC}, and $A$ is related to $\rho_{00}$ by Eq.~\eqref{eq:Arho_00}. This function can be used to extract $\rho_{00}$ with $F$ obtained from simulation. In this procedure following the usual practice we let $B=0$ and $C=0$ in Eq.~\eqref{eq:matrixABC} for the sake of simplicity, but in principle a finite $B$ value may have an effect on $\rho_{00}^{obs}$ as shown in Eq.~\eqref{eq:matrixABC}, although the effect is expected to be of secondary. The procedure with a finite $B$ included is more complicated but can be based on the same principle laid out in this paper.

Note that $F$ needs to be extracted in the same kinematic range for which the raw $\frac{d\mathrm{N}}{d\cos\theta'^* }$ was studied. Indeed the value of $F$ is dominated by the kinematic cuts thus it should be obtained for each [$p_T$,$\eta$] bin separately.

\begin{figure}[h]       
\centering
\makebox[1cm]{\includegraphics[width=0.3 \textwidth]{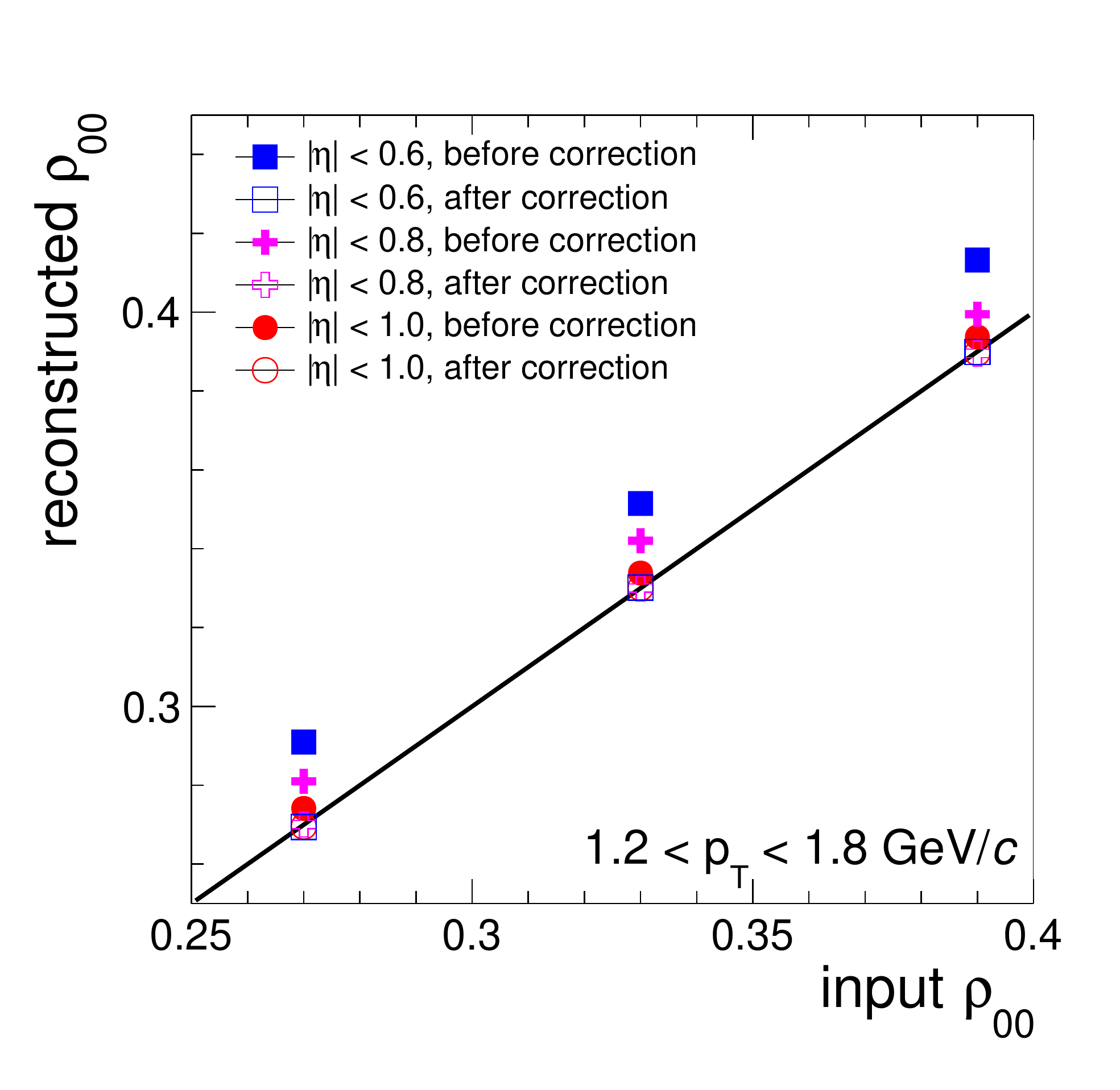}}
\caption{Simulation of the reconstructed $\rho_{00}$ versus input ones for a specific $p_T$ range [1.2 - 1.8 GeV/$c$], with (open symbols) and without (solid symbols) acceptance correction on $\eta$ coverage. Three cases are simulated for different $\eta$ coverages.
}
  \label{fig:acceptanceSim}
\end{figure}

To test the procedure, in Fig.~\ref{fig:acceptanceSim} we simulate $\phi$ particles with a characteristic $p_T$ distribution~\cite{Abelev:2007}, for a given $p_T$ range [1.2 - 1.8 GeV/$c$], and a uniform distribution in rapidity from -6 to 6. A $v_2({p_T})$ dependence is also generated according to~\cite{Adamczyk:2015fum}.  $\phi$ particles are decayed through Pythia~\cite{Sjostrand:2007gs}. A uniform distribution ($\rho_{00}=1/3$) is firstly generated, and three cuts  ($|\eta|<0.6, 0.8, 1.0$) on pseudorapidity are applied on $K^+$ and the corresponding $F$ value is obtained for each cut. Different input values of $\rho_{00}$ are achieved by selectively tossing $\phi$ particles according to $\mathrm{cos}\theta^*$ value of their daughter $K^+$, and the $\rho_{00}$ is reconstructed with (open symbols) and without (solid symbols) correction for all three sets of cut. We see a tight $\eta$ cut tends to artificially increase the observed $\rho_{00}$, which is the same observation as reported in~\cite{Lan:2017nye}. Nevertheless, with correction the reconstructed $\rho_{00}$ converge on to the right value represented by the solid line. 

\section{ Measuring spin alignment as a function of azimuthal angle} \label{sec:azimuthalDependence}

It is expected that the initial twist generated by spectators is maximum in the RP, and so is the vorticity~\cite{Becattini:2013vja}. Due to the low viscosity of the system, the vorticity may not propagate efficiently from  in- to out-of-RP, and this may lead to larger in-plane than out-of-plane global spin alignment (as well global polarization). Thus the azimuthal dependence of spin alignment carries information of the system properties and is an interesting measurement. 

One of the complications in the study of $(\Phi-\Psi)$ dependence of spin alignment arises from the correction of EP resolution. Here $\Phi$ is the azimuthal angle of $\phi$ particle in the laboratory frame. When EP is different from RP, both $\theta^*$ and $(\Phi-\Psi)$ are affected. The smearing of RP not only reduces the observed overall spin alignment, but also makes its observed azimuthal dependence weaker than what it actually is. When $\rho_{00}$ is presented as a function of $\Phi-\Psi$, $\rho_{00}$ at various $(\Phi-\Psi)$ cannot be corrected by a single resolution. That is because that particles in a given $(\Phi-\Psi_{EP})$ bin can come from every original  $(\Phi-\Psi)$ bin, and for particles that stay in the vicinity bins after RP smearing, their events experience less EP perturbation than those that end further from the original bin. 

To address this problem, we define a particle-level resolution as : 
\begin{eqnarray}
r_{ij} = \frac{\sum_{k} (m^{k}_{ij}) \cdot w_{ij}^k \cdot \mathrm{cos}[2(\Psi_{EP}^{k} - \Psi_{RP})]} {M_{ij}} ,
\end{eqnarray}
where $M_{ij}= \sum_{k} m^{k}_{ij}$, and it is particle yield from bin i before smearing ended in bin j after smearing. $m^{k}_{ij}$ is the same but for $k^{th}$ event. $w_{ij}^k = \frac{<\sum_{j}m_{ij}>}{\sum_{j}m_{ij}^{k}}$ , and $\left\langle\dots\right\rangle$ denotes the average over events. The relationship between $\rho^{obs}_{00}$ and $\rho^{real}_{00}$ can be written as

\begin{eqnarray}
\hspace{-6mm}
 \begin{pmatrix}
  \rho^{obs}_{00,1} \!-\! \frac{1}{3} \\
  \rho^{obs}_{00,2} \!-\! \frac{1}{3} \\
  \vdots                      \\
  \rho^{obs}_{00,n} \!-\! \frac{1}{3} 
 \end{pmatrix}
 \! = \!	
  \begin{pmatrix}
   a_{11}  & a_{12}  & \dots  & a_{1n} \\
   a_{21}  & a_{22}  & \dots  & a_{2n} \\
   \vdots  & \vdots  & \ddots & \vdots \\
   a_{n1}  & a_{n2}  & \dots  & a_{nn}
  \end{pmatrix}
 \begin{pmatrix}
  \rho^{real}_{00,1} \!-\! \frac{1}{3} \\
  \rho^{real}_{00,2} \!-\! \frac{1}{3} \\
  \vdots                       \\
  \rho^{real}_{00,n} \!-\! \frac{1}{3} 
  \end{pmatrix}
\end{eqnarray}
which is abridged as
\begin{eqnarray}\label{eq:rho00_real_to_obs}
\Big[ \rho_{00}^{obs} -1/3 \Big] = A \times \Big [ \rho_{00}^{real}-1/3 \Big ] .
\end{eqnarray}
The elements of this matrix is  
\begin{eqnarray}
a_{ij} = \frac{M_{ji}}{\sum_{j}{M_{ji}}} \cdot \frac{4}{1+3r_{ji}} ,
\end{eqnarray}
which takes into consideration both the yield with smearing and particle level resolution.
By taking the inverse of matrix $A$, $\rho_{00}^{real}$ can be obtained 
\begin{eqnarray}\label{eq:rho00_obs_to_real}
\Big [ \rho_{00}^{real} -1/3 \Big ] = A^{-1} \times \Big [ \rho_{00}^{obs}-1/3 \Big ] .
\end{eqnarray}

In practice, we know that $\phi$-meson's azimuthal angle distribution is described by
\begin{eqnarray}
\frac{dN}{d(\Phi-\Psi)} \propto 1+2v_{2}\mathrm{cos}2(\Phi-\Psi) ,
\label{eq:v2_phi_dis}
\end{eqnarray}
and the perturbation of RP, $\Delta$, can be generated by the probability density function of Eq.~\eqref{eq:Detla_Psi_dis}. For event $k$ with a given EP, $m^{k}_{ij}$ and $\mathrm{cos}[2(\Psi_{EP}^{k} - \Psi_{RP})]$ can be calculated. Repeating this step for many events, $M_{ij}$ and $r_{ij}$ can be determined, and so can be the matrix element.

\begin{figure}    
\centering
\includegraphics[width=0.5 \textwidth]{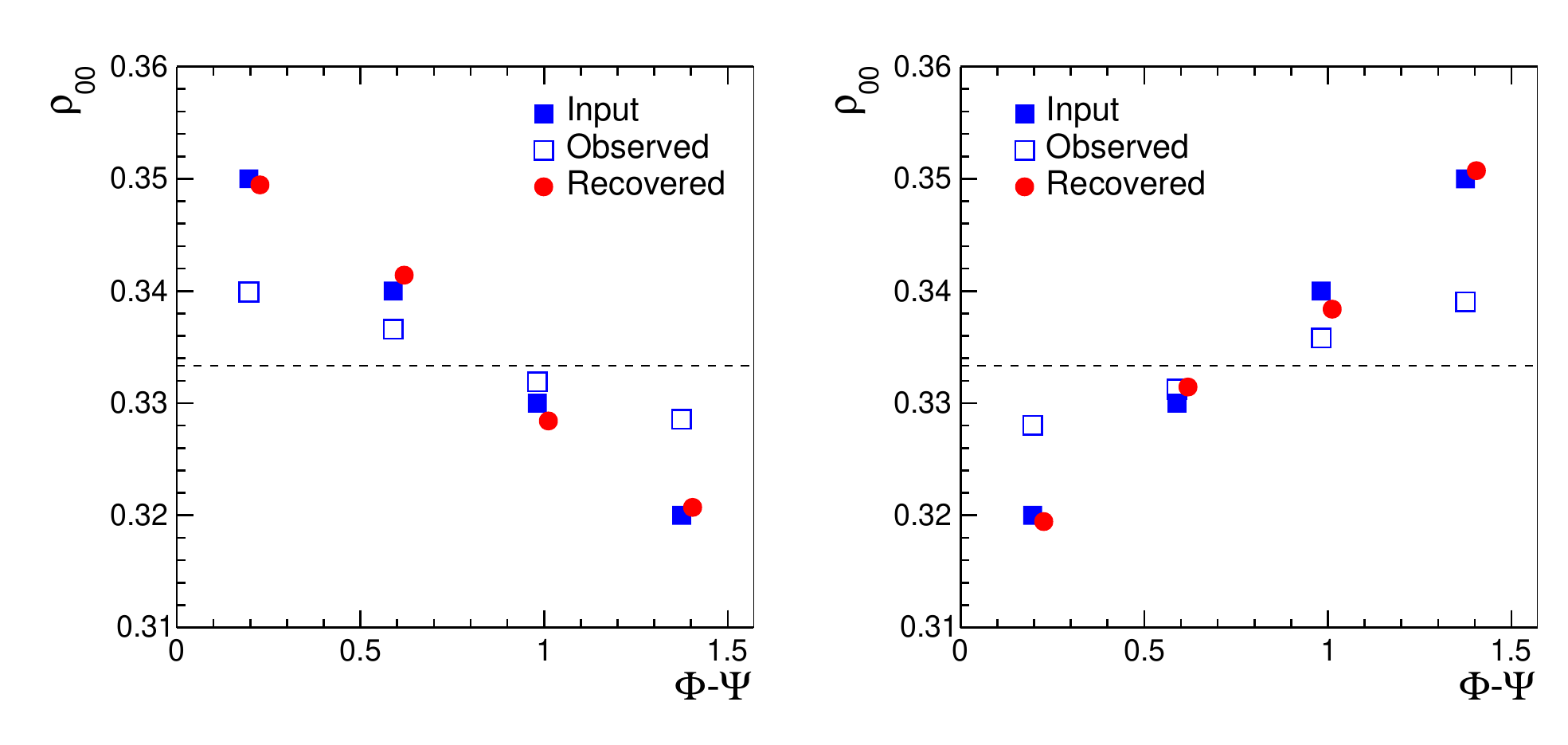}
\caption{Input, observed and recovered $\rho_{00}$ as a function of $\Phi-\Psi$, based on simulated data.}
  \label{fig:rho00_phi_simu}
\end{figure}

Fig.~\ref{fig:rho00_phi_simu} shows the correction at work with a toy model simulation. In this simulation, we assume that the azimuthal angle($\Phi$) of $\phi$-meson has no correlation with the $\theta^{*}$ of the daughter kaon in the rest frame. A set of $\mathrm{cos}\theta^{*}$ distributions were generated with Eq.~\eqref{eq:v2_phi_dis}, with different ($\Phi-\Psi$) bins taking different $\rho_{00}$ values. An azimuthal angel distribution is also generated with $v_{2} = 0.08$ according to Eq.~\eqref{eq:v2_phi_dis}. $\Delta$ for $\Psi_{EP}$ is generated with a probability density function of Eq.~\eqref{eq:Detla_Psi_dis} with $\chi=0.69$. We see that, due to EP smearing, the observed $\rho_{00}$ has weaker ($\Phi-\Psi$) dependence than what the input one has, however the recovered $\rho_{00}$ values agree well with the corresponding input. 

Note that this procedure can also be used to correct the azimuthal angle dependence of $\Lambda(\bar{\Lambda})$ global polarization ($P_{H}$). To do that, one simply needs to replace ($\rho_{00}-1/3$) by $P_H$, and replace the elements of matrix $A$ by
\begin{eqnarray}
a_{ij} = \frac{M_{ji}}{\sum_{j}{M_{ji}}} \cdot {r_{ji}} \, ,
\end{eqnarray}
and $r_{ij}$ by
\begin{eqnarray}
r_{ij} = \frac{\sum_{k} (m^{k}_{ij}) \cdot w_{ij}^k \cdot \mathrm{cos}[n(\Psi_{EP,n}^{k} - \Psi_{RP})]} {M_{ij}},
\end{eqnarray}
where $n$ denotes the harmonic for EP.

The study of azimuthal dependence of global spin alignment is also complicated by local spin alignment, namely, the finite spin alignment in the center-of-mass helicity (HX) frame.  Such apparent spin alignment can be caused either by physics or by finite detector efficiency.  In the HX frame, the quantization $z$-axis is chosen to be the momentum direction of parent (in our case, $\phi$ particle). Because so, the global spin alignment will have a trivial $(\Phi-\Psi)$ dependence. This can be understood with an example of extreme spin alignment in HX frame : if $K^+$ are produced in the exact same or opposite direction of $\phi$-particle's momentum,  then for $\phi$-particles produced in- (out-of-) RP, their daughter $K^+$are found completely in the in- (out-of) RP direction, which gives rise to a trivial  $(\Phi-\Psi)$ dependence of global spin alignment. In fact, this effect will be there w.r.t any given plane, not just for RP.  In addition, a finite acceptance in pseudorapidity could also introduce a trivial  $(\Phi-\Psi)$ dependence of global spin alignment. This can be understood as ball-shape objects being slightly trimmed at different azimuthal angles according to the ball's own azimuthal angle, and what is left is a shape that has an azimuthal dependence. Due to the complicated nature of the problem, these effects will be studied in a future work.

\section{ Conclusion} \label{sec:conclusion}
The measurement of global spin alignment of vector mesons is complicated by many factors, and we have provided procedures to handle a few major ones of them. We have derived a formula for correction for event plane resolution, and shown that that formula offers insights in describing $\rho_{00}$ measured with different event planes. The same formula also tells us that one should not expect zero signal with event plane with random angle in azimuth. We have presented a procedure to correct for the effect of finite pseudorapidity coverage, which can be used in experimental setups with cylindrical symmetry.  When presenting the azimuthal angle dependence, the result cannot be corrected for finite event plane resolution with a single value, and we have presented a way to correct for it thoroughly. The knowledge presented in this paper allows the measurement of global spin alignment to be conducted properly and accessible differentially. 


\section*{Acknowledgements}
We'd like to thank S. Voloshin, G. Wang and X. Sun for fruitful discussions. B. Tu is supported in part by MoST of China 973-Project No. 2015CB856901, NSFC under grant No. 11521064. C.S. Zhou is supported in part by the National Natural Science Foundation of China under Contracts No. 11421505 and No. 11220101005, the Major State Basic Research Development Program in China under Contract No. 2014CB845401. A.H.T. is supported by the U.S. Department of Energy, Office of Science, under Grant DE-SC0012704.

{}


\end{document}